\documentstyle[aps,prb,eqsecnum,epsfig,here]{revtex}
\begin{document}
\title{High field magnetotransport in composite conductors:\\
The effective medium approximation revisited}

\author{David J. Bergman$^{a,b}$ and David G. Stroud$^b$}
\address{$^a$School of
Physics and Astronomy, Raymond and Beverly Sackler Faculty of Exact
Sciences\\Tel Aviv University,
Tel Aviv 69978, Israel\\
$^b$Department of Physics, The Ohio State University,
Columbus, OH 43210-1106}

\date{\today}

\maketitle

\def\be{\begin{equation}}
\def\ee{\end{equation}}

\widetext

\begin{abstract}

The self consistent effective medium approximation (SEMA)
is used to study three-dimensional random conducting composites 
under the influence of a strong magnetic field {\bf B}, in the
case where all constituents exhibit isotropic response.
Asymptotic analysis is used to obtain almost closed form results
for the strong field magnetoresistance and Hall resistance
in various types of two- and three-constituent isotropic mixtures
for the entire range of compositions. Numerical solutions of the 
SEMA equations are also obtained, in some cases, and compared with 
those results. In two-constituent free-electron-metal/perfect-insulator
mixtures, the magnetoresistance is asymptotically proportional
to $|{\bf B}|$ at {\em all concentrations above the percolation
threshold}. In three-constituent
metal/insulator/superconductor mixtures a line of critical
points is found, where the strong field magnetoresistance switches
abruptly from saturating to non-saturating dependence on $|{\bf B}|$,
at a certain value of the insulator-to-superconductor concentration
ratio. This transition appears to be related to the phenomenon
of anisotropic percolation.

\end{abstract}


\vspace{-0.5 cm}
\noindent
To appear in Phys.\ Rev.\ {\bf B}


\vspace{-0.5 cm}

\section{Introduction}
\label{introduction}

Over the years, the Bruggeman self consistent effective medium
approximation (SEMA) for the bulk effective electrical conductivity
of a composite medium \cite{Bruggeman35,LandauerJAP52} has spawned
a number of extensions and generalizations. Those include a SEMA
for elastic stiffness of a composite medium, \cite{Budiansky,Hill}
a SEMA for a weakly nonlinear conducting composite, \cite{BergmanPRB89}
and a SEMA for a strongly nonlinear (power law) conducting composite.
\cite{BergmanTrieste91,SaliBergman97} They also include a SEMA for a linear
conducting composite medium where the constituents are characterized
by {\em nonscalar conductivity tensors}. \cite{Stachowiak70,StroudPRB75}
Such approximations are needed to study  the nonscalar conductivity
due to an externally applied magnetic field {\bf B}.
\cite{Stachowiak70,StroudPRB75,CohenJortner73,StroudPanPRB76,Balagurov86}

In composite conductors, the microstructure can induce a dependence
of the Ohmic resistivity on magnetic field (i.e., magnetoresistance) even
when none of the constituents exhibits any such dependence by
itself. This arises because the local current density ${\bf J}({\bf r})$
acquires spatial fluctuations in both magnitude and direction,
as a result of the heterogeneity or microstructure.
Those are reflected in similar fluctuations of the local Hall
(electric) field ${\bf E}_H({\bf r})$, which usually has a nonzero
component along the direction of the volume averaged current
density $\langle{\bf J}\rangle$. The volume average of that
component of ${\bf E}_H$ vanishes to leading order in {\bf B},
but in higher orders that average is usually nonzero.  As a result,
the bulk effective Ohmic resistivity of the
composite usually depends upon {\bf B}.
The SEMA developed in Refs.\ \onlinecite{Stachowiak70,StroudPRB75}
was used in numerical studies of magnetotransport in various types of
composites, where this induced magnetoresistance
played an important role.
\cite{StroudPRB75,StroudPanPRB76,StroudPanPRB79,StroudPRL80}
It was also used in an unpublished discussion of asymptotic strong
field behavior of magnetoresistance in a free-electron-conductor host
with open orbit inclusions. \cite{PanThesis}

More recently, it was shown that this type of SEMA, when 
applied to composites with a {\em columnar
microstructure}, often violates some exact relations
which exist between the different components of the bulk
effective resistivity tensor.  A modified
SEMA has been developed which incorporates those exact relations
[the ``columnar unambiguous self-consistent effective medium
approximation (CUSEMA)'']; it has been used for a detailed
asymptotic analysis of metal/insulator $(M/I)$ and
metal/superconductor $(M/S)$ random columnar mixtures in the 
strong field limit. \cite{CUSEMA} Very recently, the CUSEMA
was applied to a three-constituent metal/insulator/superconductor
$(M/I/S)$ random columnar composite mixture;
a line of critical points was found to appear whenever the $I$ and the $S$
constituents are present in equal amounts, i.e., when their
volume fractions $p_I$ and $p_S$ are equal. At this point,
the asymptotic strong field behavior of the magnetoresistance
switches abruptly from saturated for $p_S>p_I$ to nonsaturating
$\propto|{\bf B}|^2$ behavior for $p_S<p_I$. \cite{BergmanPRL2000}
This critical point exhibits scaling behavior, along with
critical exponents and scaling functions, all of which were
determined, approximately, using CUSEMA.

In this article we present a theoretical study of magnetoresistance
in three-dimensional disordered composite media. Some of the systems
we study are isotropic, two-constituent mixtures of isotropic
constituents.  In this category, we consider
mixtures of normal conductors $(M_1/M_2)$, $M/I$ mixtures, and 
$M/S$ mixtures. 
No ``intrinsic
anisotropy'' is allowed for any constituent; thus, open orbit
constituents are not considered. The normal conducting constituents
can be simple free-electron conductors, in which case only the
Hall resistivity depends upon {\bf B}, or they can be more
complicated conductors or semiconductors where the Ohmic
resistivity also depends upon {\bf B}. But the longitudinal
Ohmic resistivity $\rho_\parallel$ (along {\bf B}) and the
transverse Ohmic resistivity $\rho_\perp$ (perpendicular to
{\bf B}), as well as the Hall resistivity $\rho_{\rm Hall}$,
are assumed to be {\em independent of the direction} of {\bf B}.
We also consider an $M/I/S$ mixture, again with isotropic
constituents, an isotropic microstructure, and 
$\rho^{(M)}_\parallel$, $\rho^{(M)}_\perp$, $\rho^{(M)}_{\rm Hall}$ 
of the $M$ constituent which are independent
of the direction of {\bf B}. 

The SEMA equations
for these composites do not admit exact closed form solutions.
However, in the strong field limit, asymptotic analysis can
be applied to those equations.
The strong field limit means that $|H| \gg 1$, where $H$ is the
Hall-to-transverse-Ohmic resistivity ratio in a metallic or
normal conducting constituent
\be
H\equiv\frac{\rho^{(M)}_{\rm Hall}}{\rho^{(M)}_\perp}=\mu|{\bf B}|,
\label{H_def}
\ee
and where $\mu$ is the Hall mobility; note that 
$\mu$ and $\rho^{(M)}_{\rm Hall}$, and therefore also $H$, can be
either positive or negative, depending on the sign of the
majority charge carriers.
This asymptotic analysis often leads to results in
simple closed form, or at least ``almost closed form.''
In several cases we have also solved the SEMA equations numerically
in order to compare with the asymptotic a\-na\-ly\-sis.

The rest of this article is organized as follows.
In Section \ref{review} we briefly review the SEMA, using a
formulation that reproduces the usual SEMA equations for
non-scalar conducting constituents in terms 
of depolarization coefficients of
the different types of inclusions. In Sections
\ref{two_constituents} and \ref{MIS}
asymptotic a\-na\-ly\-sis is used to find almost closed
form solutions to those equations for a number of
special cases. Those asymptotic solutions are compared,
in some cases, with numerical solutions of the same
equations. In Section \ref{two_constituents}, we treat the special
case of an $M_1/M_2$ binary mixture.
All concentrations are
considered, starting from the limit of extreme dilution and all the way to
the percolation threshold. 
In Section \ref{MIS},
$M/I/S$ three-constituent mixtures
are treated using asymptotic a\-na\-ly\-sis. A line of critical
points is found where the strong field response changes
abruptly in the manner of a second-order-phase-transition,
as the relative amounts of $I$ and $S$ are varied.
Critical exponents, a scaling variable, and scaling functions
are found which characterize the critical behavior.
We also discuss $M/S$ and $M/I$ mixtures as special cases of
these three-constituent composites.  A linear dependence of the
magnetoresistance upon $|{\bf B}|$ is found in 
free-electron-metal/perfect-insulator mixtures for {\em all
concentrations} where the mixture conducts macroscopically. 

Section \ref{summary} provides a discussion of the
results. We formulate a physical picture of the
microscopic processes (i.e., local current flow patterns)
which lead to some of the results
found in previous sections for the macroscopic response.
We also discuss a possible relation between the line of critical
points found in the macroscopic magnetotransport of
three-constituent $M/I/S$ mixtures and the phenomenon of
anisotropic percolation.

\section{Review of SEMA for the current problems}
\label{review}

The self consistent effective medium approximation (SEMA) for
constituents with arbitrary conductivity tensors, which appear
in the system as ellipsoidal grains, was 
developed many years ago by one of the present authors
\cite{StroudPRB75}. This development followed more specialized versions,
such as all scalar conductivities,
\cite{Bruggeman35,LandauerJAP52} and later,
strong field magnetotransport \cite{Stachowiak70} and 
weak field Hall conductivity. \cite{CohenJortner73}
Here we describe a slightly different formulation of the
general theory of Ref.\ \onlinecite{StroudPRB75}.  We use this formulation
to find asymptotic physical 
solutions of the SEMA equations when the magnetic field is very
strong. This formulation was described in detail in Ref.\
\onlinecite{CUSEMA}.
A similar formulation also appeared in Ref.\ \onlinecite{Balagurov86}.

In practice, the SEMA requires that one calculate the 
electric field ${\bf E}_1$ induced inside a single inclusion,
with conductivity tensor $\hat\sigma_{\rm inc}$, embedded in
an otherwise uniform host with conductivity tensor $\hat\sigma_{\rm host}$,
when an external uniform electric field ${\bf E}_0$ is applied
at large distances. Whenever the inclusion is an ellipsoid,
${\bf E}_1$ is {\em uniform}, whatever the values of
$\hat\sigma_{\rm inc}$, $\hat\sigma_{\rm host}$. Obviously,
${\bf E}_1$ will be a homogeneous linear function of ${\bf E}_0$,
which can be written with the help of a matrix
$\hat\gamma_{\rm inc}(\hat\sigma_{\rm inc},\hat\sigma_{\rm host})$
\be
{\bf E}_1=\hat\gamma_{\rm inc}\cdot{\bf E}_0.
\ee
If the coordinate axes are taken to lie
along the principal axes of the symmetric part of the tensor
$\hat\sigma_{\rm host}$, then we can write
\begin{eqnarray}
\left(\frac{1}{\hat\gamma_{\rm inc}}\right)_{\alpha\gamma}&=&
\delta_{\alpha\gamma}-\sum_\beta
\frac{n_{\alpha\beta}\delta\sigma_{\beta\gamma}}
{\left(\sigma^{({\rm host})}_{\alpha\alpha}
\sigma^{({\rm host})}_{\beta\beta}\right)^{1/2}},\label{gamma_inverse}\\
\delta\hat\sigma&\equiv&\hat\sigma_{\rm host}-
\hat\sigma_{\rm inc}.
\label{delta_sigma}
\end{eqnarray}
The factors $n_{\alpha\beta}$ are elements of the depolarization
tensor $\hat n$, {\em not of the actual physical inclusion}, but of its
image after the coordinate axes have been rescaled by the following
transformation
\be
x'\equiv\frac{x}{\sqrt{\sigma^{({\rm host})}_{xx}}},\;\;\;
y'\equiv\frac{y}{\sqrt{\sigma^{({\rm host})}_{yy}}},\;\;\;
z'\equiv\frac{z}{\sqrt{\sigma^{({\rm host})}_{zz}}}.
\label{axes_rescaling}
\ee
Under this transformation, the ellipsoidal inclusion is
usually transformed into another ellipsoidal shape,
with major axes that have different lengths and different
orientations compared to those of the physical inclusion.
The SEMA is then obtained by setting $\hat\sigma_{\rm host}=\hat\sigma_e$,
and demanding that the small but exactly calculable
change in $\langle{\bf J}\rangle$, caused
by one isolated ellipsoidal inclusion, vanish 
when averaged over the different kinds of inclusions. This leads 
to the following self consistency equations for the elements
of $\hat\sigma_e$
\be
0=\left\langle\left(\hat\sigma_e-\hat\sigma_{\rm inc}
\right)\cdot\hat\gamma_{\rm inc}
(\hat\sigma_{\rm inc},\hat\sigma_e)\right\rangle.
\label{naive_EMA}
\ee
In general, these are nonlinear equations, in which there appear
{\em non-elementary transcendental functions} of the elements
of $\hat\sigma_e$. Thus, closed form solutions are usually
out of the question. However, we will show below that asymptotic
solutions, for a very strong magnetic field {\bf B}, can
sometimes be obtained in almost closed form.  For smaller values
of {\bf B}, the SEMA equations can be solved numerically, where
needed; these solutions are discussed below.

We assume the following forms for the resistivity matrices of the
host and inclusion:
\be
\hat\rho_{\rm host}=\rho_0\left(\begin{array}{ccc}
\alpha & -\beta & 0 \\
\beta & \alpha & 0 \\
0 & 0 & \lambda\end{array}\right),\;\;\;
\hat\rho_{\rm inc}=\rho_1\left(\begin{array}{ccc}
1 & -H & 0 \\
H & 1 & 0 \\
0 & 0 & \nu\end{array}\right).\label{rho_host_const}
\ee
These forms mean that both the host and the inclusion have
isotropic electrical response, and that the only physically
selected direction is that of the magnetic field {\bf B},
which is always taken to lie along $z$. This implies that
the microstructure, as well as the electrical response of
all the constituents, are isotropic. The $\alpha$ coefficient
in $\hat\rho_{\rm host}$ is actually redundant. We will
sometimes exploit that redundancy by choosing $\rho_0$
of the bulk effective resistivity tensor of the composite
medium to be the
same as one of the constituent values of $\rho_1$. The
conductivity tensors $\hat\sigma_{\rm host}$, $\hat\sigma_{\rm inc}$
are easily found by inverting $\hat\rho_{\rm host}$,
$\hat\rho_{\rm inc}$. 

Because we are considering isotropic 
microstructures, we will assume that all the constituents
appear as spherical inclusions in the fictitious uniform
effective medium host. In that case we find
\widetext
\be
\hat\gamma_{\rm inc}(\hat\sigma_{\rm inc},\hat\sigma_{\rm host})
=\left(\begin{array}{ccc}
\frac{1}{D_{\rm inc}}\left(1-n_x+n_x\frac{\rho_0}{\rho_1}
\frac{\alpha^2+\beta^2}{\alpha(1+H^2)}\right) &
\frac{n_x}{D_{\rm inc}}\left(\frac{\beta}{\alpha}-
\frac{\rho_0}{\rho_1}\frac{\alpha^2+\beta^2}{\alpha}
\frac{H}{1+H^2}\right) & 0 \\
-\frac{n_x}{D_{\rm inc}}\left(\frac{\beta}{\alpha}-
\frac{\rho_0}{\rho_1}\frac{\alpha^2+\beta^2}{\alpha}
\frac{H}{1+H^2}\right) & 
\frac{1}{D_{\rm inc}}\left(1-n_x+n_x\frac{\rho_0}{\rho_1}
\frac{\alpha^2+\beta^2}{\alpha(1+H^2)}\right) & 0 \\
0 & 0 & \frac{1}{1-n_z+n_z\lambda/\nu}\end{array}\right),
\label{gamma_inc}
\ee
where
\be
D_{\rm inc}\equiv\left(1-n_x+n_x\frac{\rho_0}{\rho_1}
\frac{\alpha^2+\beta^2}{\alpha(1+H^2)}\right)^2 +
n_x^2\left(\frac{\beta}{\alpha}-
\frac{\rho_0}{\rho_1}\frac{\alpha^2+\beta^2}{\alpha}
\frac{H}{1+H^2}\right)^2,
\label{D_inc}
\ee
\noindent
and where $n_x$, $n_y=n_x$, and $n_z$ are the depolarization factors of
the spheroidal shape into which the spherical inclusion was
transformed by Eqs.\ (\ref{axes_rescaling}). Those depolarization
factors are elementary transcendental functions of
$\alpha$, $\beta$, $\lambda$. \cite{LanLif}
Thus, the elements of $\hat\gamma_{\rm inc}$ are also non-algebraic
functions of those parameters. Consequently, the equations
obtained for those parameters, by applying the self consistency
requirements implied by Eq.\ (\ref{naive_EMA}), will usually
be complicated, coupled, non-algebraic equations which lack
closed form solutions in terms of elementary functions. This
makes the qualitative study of their physical solutions highly
nontrivial.

\section{Application of SEMA to magnetoresistance of an
$M_1/M_2$ mixture}
\label{two_constituents}

As an illustration, we now work out the effective resistivity tensor for
a binary composite of two normal metals, using the SEMA.  Two other
special cases---a composite of normal metal and insulator, and one of
normal metal and perfect conductor---will be discussed in the next section
as special cases of three-constituent mixtures.

In the present case, the bulk effective resistivity tensors of the two constituents
are assumed to have the similar forms
\be
\hat\rho_1=\rho_1\left(\begin{array}{ccc}
1 & -H_1 & 0 \\
H_1 & 1 & 0 \\
0 & 0 & \nu_1\end{array}\right),\;\;\;
\hat\rho_2=\rho_2\left(\begin{array}{ccc}
1 & -H_2 & 0 \\
H_2 & 1 & 0 \\
0 & 0 & \nu_2\end{array}\right),
\label{rho_1_rho_2}
\ee
while the bulk effective resistivity tensor is assumed to
have the form
\be
\hat\rho_e=\rho_1\left(\begin{array}{ccc}
\alpha & -\beta & 0 \\
\beta & \alpha & 0 \\
0 & 0 & \lambda\end{array}\right).
\label{rho_e}
\ee
Note that $\hat\rho_e$ and $\hat\rho_1$ have been expressed in
terms of the same resistivity factor
$\rho_1$.  The forms (3.1) and (3.2) imply
that both constituents are isotropic conductors, as is the
composite, and that the magnetic field {\bf B} is applied along
$z$. They include the case where both $\hat\rho_1$ and $\hat\rho_2$
represent simple free-electron or free-hole conductors.  They also
include more general types of conductors, where the Ohmic
resistivities and Hall mobility depend upon $|{\bf B}|$.

Application of Eq.\ (\ref{naive_EMA}) leads to three coupled
equations for the unknown parameters $\alpha$, $\beta$, $\lambda$,
which arise from the $zz$, $xy$, and $xx$ components of that tensorial
equation [the $xx$ and $yy$ components lead to the same equation,
as do the $xy$ and $yx$ components, while the $xz$, $zx$, $yz$, and
$zy$ components of Eq.\ (\ref{naive_EMA}) vanish identically;
$p_1$ and $p_2=1-p_1$ denote the volume fractions of the two
constituents]:
\begin{eqnarray}
0&=&\frac{1-n_z}{\lambda}
-\frac{\rho_1}{\rho_2}\frac{\lambda n_z}{\nu_1\nu_2}
+n_z\left(\frac{p_2}{\nu_1}+\frac{p_1}{\nu_2}\frac{\rho_1}{\rho_2}\right)
\nonumber\\
&&\;-\;(1-n_z)\left(\frac{p_2}{\nu_2}\frac{\rho_1}{\rho_2}
+\frac{p_1}{\nu_1}\right),\label{zz_MM}\\
0&=&\left(p_2\frac{D_1}{D_2}+p_1\right)\frac{\beta}{\alpha^2+\beta^2}
-p_2\frac{D_1}{D_2}\frac{\rho_1}{\rho_2}\frac{H_2}{1+H_2^2}
\nonumber\\
&&\;-\;p_1\frac{H_1}{1+H_1^2},\label{xy_MM}\\
0&=&\left(p_2\frac{D_1}{D_2}+p_1\right)
\left(\frac{\alpha}{\alpha^2+\beta^2}-\frac{n_x}{\alpha}\right)
\nonumber\\
\lefteqn{
\;+\;\frac{p_2}{1+H_2^2}\frac{D_1}{D_2}\frac{\rho_1}{\rho_2}
\left[n_x\left(2+\frac{2\beta H_2}{\alpha}-\frac{\rho_1}{\rho_2}
\frac{\alpha^2+\beta^2}{\alpha}\right)-1\right]}\nonumber\\
&&\;+\;\frac{p_1}{1+H_1^2}\left[n_x\left(2+\frac{2\beta H_1}{\alpha}
-\frac{\alpha^2+\beta^2}{\alpha}\right)-1\right].\label{xx_MM}
\end{eqnarray}
Not surprisingly, these equations do not admit any closed form solutions. Since
we are interested in the behavior of $\alpha$, $\beta$, $\lambda$
at strong fields, we will apply asymptotic a\-na\-ly\-sis for the limiting
case where $H_1\equiv H$ and $H_2\equiv\omega H$ are both much greater 
than 1, i.e., $\omega$ remains bounded but $|H|\gg 1$. All other
physical parameters of the two constituents are also assumed to 
remain bounded. From Eq.\ (\ref{zz_MM}) it follows that $\lambda$
must also remain bounded. We have tried various ansatzes for the
asymptotic behavior of $\alpha$, $\beta$, and proceeded to
examine whether they lead to a mathematically consistent and
physically admissible solution. The only ansatz that could satisfy
those requirements was
\be
\beta\cong\beta_0 H,\;\;\;\alpha\cong\alpha_0|H|^{2/3},\;\;\;
\lambda\cong\lambda_0,\label{ansatz_MM}
\ee
where $\alpha_0$, $\beta_0$, $\lambda_0$ are positive dimensionless
coefficients. It follows from this ansatz that the transformation of
Eqs.\ (\ref{axes_rescaling}) changes the spherical inclusions into
flat, pancake shaped or oblate spheroids, with the short major
axis along $z$, and with eccentricity $e$ that diverges as 
$|H|\rightarrow\infty$. The eccentricity $e$ and depolarization factors
are given by \cite{LanLif}
\begin{eqnarray}
e&=&\left(\frac{\alpha^2+\beta^2}{\alpha\lambda}-1\right)^{1/2}
\cong\frac{\beta_0}{\sqrt{\alpha_0\lambda_0}}|H|^{2/3}\gg 1,
\label{e_MM}\\
n_z&=&\frac{1+e^2}{e^3}(e-\arctan e)\cong 1-\frac{\pi}{2e}\label{nz_MM}\\
&\cong&
1-\frac{\pi}{2}\frac{\sqrt{\alpha_0\lambda_0}}{\beta_0}|H|^{-2/3}
\cong 1,\label{nz_MM_1}\\
n_x&=&n_y=\frac{1}{2}(1-n_z)=
\frac{\pi}{4}\frac{\sqrt{\alpha_0\lambda_0}}{\beta_0}|H|^{-2/3}
\ll 1.\label{nx_MM}
\end{eqnarray}
This leads to $D_1\cong 1$ and $D_2\cong 1$, and when these results
are used in Eqs.\ (\ref{xy_MM}) and (\ref{xx_MM}), one arrives
at the following explicit asymptotic expressions for 
$\alpha_0$, $\beta_0$, $\lambda_0$:
\begin{eqnarray}
\beta_0&=&\frac{1}{p_1+p_2\frac{H_1\rho_1}{H_2\rho_2}},\label{beta_MM}\\
\lambda_0&=&p_1\nu_1+\frac{\rho_2}{\rho_1}p_2\nu_2,\label{lambda_MM}\\
\alpha_0&=&{\left(\frac{\pi}{4}p_1p_2\right)^{2/3}
\left|1-\frac{H_1\rho_1}{H_2\rho_2}\right|^{4/3}
\left(p_1\nu_1+\frac{\rho_2}{\rho_1}p_2\nu_2\right)^{1/3}
\over\left(p_1+p_2\frac{H_1\rho_1}{H_2\rho_2}\right)^2}.\nonumber\\
&&\label{alpha_MM}
\end{eqnarray}
Note that $\alpha_0=0$ when the Hall
resistivities of the two constituents are equal
$H_1\rho_1=H_2\rho_2$.  Moreover, when
$\rho_2/\rho_1\gg 1$, then $\beta_0$ tends to the finite value
$1/p_1$, while $\lambda_0$ and $\alpha_0$ both diverge.
As we shall see in Section \ref{MI} below, this be\-ha\-vi\-or presages
the behavior exhibited by those parameters when the No.\ 2
constituent is a perfect insulator, i.e., when $\rho_2=\infty$.

A similar asymptotic dependence of the
transverse relative bulk effective magnetoresistance
on magnetic field was found previously for
a free-electron-metal host with open orbit inclusions. \cite{PanThesis} 
A similar asymptotic
dependence was also previously found by Dreizin and Dykhne,
who presented a qualitative microscopic discussion of
magnetotransport in composite media. \cite{DykhneJETP73}
In our language, those results would translate into
$\alpha\propto|H|^{2/3}$, as we have also found. While
the asymptotic behavior found in those previous studies
is similar to what we find here, the present discussion
shows that (a) one does not need to have open orbits in any
constituent in order to observe this kind of behavior,
(b) the behavior we find is a straightforward consequence
of SEMA. Since SEMA becomes exact in the dilute limit,
when either $p_1$ or $p_2$ is very small, we believe that the
asymptotic result $\alpha\propto|H|^{2/3}$ is exact. It appears that the
only requirement for obtaining this type of asymptotic response is
that the two constituents have comparable Ohmic resistivities
and different Hall resistivities.  In fact, our numerical solutions
of these equations also suggest that only the 
difference in the Hall resistivities
is crucial: The $|H|^{2/3}$ power law is obtained even if the
two constituents have {\em the same Ohmic resistivities}.

In order to confirm that our ansatz for the asymptotic behavior is indeed
correct, we have solved the SEMA equations numerically for a simple
example.  We assume that $\rho_1 = \rho_2$, $\nu_1=\nu_2=1$,
and $H\equiv H_1 = 2H_2$, and we consider $p_1 = p_2 = 0.5$.  In
Fig.\ \ref{magnetocond}(a) we plot the relative
bulk effective transverse magnetoresistivity $\alpha - 1$
vs.\ $|H|^{2/3}$. It is evident that this quantity
rapidly approaches a linear dependence on
$|H|^{2/3}$ as $|H|$ increases, and that the asymptotic dependence found
analytically above is accurately achieved for $|H| > 2 $ 
for this choice of parameters.  
In Fig.\ \ref{magnetocond}(b), we plot the quotient
$\beta/H=\rho^{(e)}_{\rm Hall}/\rho^{(1)}_{\rm Hall}$ vs.\ $|H|^{2/3}$.
The second form of this quotient shows that it is equal to
the ratio of bulk effective Hall resistivity to the Hall
resistivity of the No.\ 1 constituent, suggesting the name
``relative Hall resistivity''.
Clearly, $\beta$ also rapidly approaches its
asymptotic dependence upon $H$, which is
linear rather than $\propto|H|^{2/3}$ i.e., the ratio $\beta/H$
becomes field-independent.

\section{Three-constituent $M/I/S$ mixtures}
\label{MIS}

\subsection{General considerations}

Next, we turn to a discussion of a three-constituent composite containing
a volume fraction $p_M$ of normal metal, $p_S$ of perfect conductor,
and $p_I$ of insulator.  As part of this discussion, we will consider
the special cases of two-constituent $N/S$ and $N/I$ mixtures.  We use the
subscripts $M$, $I$, and $S$ to denote normal metal, insulator, and
perfect conductor.

We assume the following forms and inequalities
for the bulk effective resistivity
tensor $\hat\rho_e$, and for the three constituent resistivity
tensors $\hat\rho_M$, $\hat\rho_I$, $\hat\rho_S$:
\begin{eqnarray}
\hat\rho_e&=&\rho_M\left(\begin{array}{ccc}
\alpha & -\beta & 0 \\
\beta & \alpha & 0 \\
0 & 0 & \lambda\end{array}\right),\;\;\;
\hat\rho_M=\rho_M\left(\begin{array}{ccc}
1 & -H & 0 \\
H & 1 & 0 \\
0 & 0 & \nu\end{array}\right),\nonumber\\
&&\label{rho_e_rho_M}\\
\hat\rho_S&=&\rho_S\hat I,\;\;\;\hat\rho_I=\rho_I\hat I,\;\;\;
\rho_S\ll\rho_M\ll|H|\rho_M\ll\rho_I,\label{rho_S_rho_I}
\end{eqnarray}
where $\hat I$ is the unit tensor. We use the results of Eqs.\
(\ref{gamma_inc}) and (\ref{D_inc}) in Eq.\ (\ref{naive_EMA}) for
this system, and then take the limits $\rho_S\rightarrow 0$ and
$\rho_I\rightarrow\infty$ to get the following equations for
$\alpha$, $\beta$, $\lambda$ from the $zz$, $xy$, and $xx$
components of that equation [again, the other components of
Eq.\ (\ref{naive_EMA}) lead either to redundancies or to identities;
$p_M$, $p_I$, $p_S$ denote the constituent volume fractions,
which satisfy $p_M+p_I+p_S=1$]
\begin{eqnarray}
\frac{\lambda}{\nu}&=&\frac{1-n_z}{n_z}\frac{n_z-p_S}{p_S+p_M-n_z},
\label{zz_MIS}\\
\frac{\beta}{\alpha^2+\beta^2}&=&
p_M\frac{H}{1+H^2}\left/\left(\frac{D_M}{D_I}p_I+p_M\right)\right.,
\label{xy_MIS}
\end{eqnarray}
\begin{eqnarray}
\lefteqn{
\left(\frac{D_M}{D_I}p_I+p_M\right)\left(\frac{\alpha}{\alpha^2+\beta^2}
-\frac{n_x}{\alpha}\right)-D_M\frac{p_S}{n_x}\frac{\alpha}{\alpha^2+\beta^2}
=}\nonumber\\&&\;\;\;\;=\;
\frac{p_M}{1+H^2}\left[1+n_x\left(\frac{\alpha^2+\beta^2}{\alpha}
-2-\frac{2\beta H}{\alpha}\right)\right],\label{xx_MIS}
\end{eqnarray}
where
\begin{eqnarray}
D_M&=&\left(1-n_x+n_x\frac{\alpha^2+\beta^2}{\alpha(1+H^2)}\right)^2
\nonumber\\
&&\;+\;\frac{n_x^2\beta^2}{\alpha^2}\left(1-\frac{\alpha^2+\beta^2}{\beta}
\frac{H}{1+H^2}\right)^2,\label{DM_MIS}\\
D_I&=&(1-n_x)^2+\frac{n_x^2\beta^2}{\alpha^2}.\label{DS_MIS}
\end{eqnarray}
Combining Eqs.\ (\ref{xy_MIS}) and (\ref{xx_MIS}) we can simplify
Eq.\ (\ref{xx_MIS}) slightly to get
\begin{eqnarray}
\lefteqn{
1+n_x\left[\frac{\alpha^2+\beta^2}{\alpha}\left(1+\frac{H}{\beta}\right)
-2-\frac{2\beta H}{\alpha}\right]=}\nonumber\\
&&\;\;\;=\;\frac{\alpha H}{\beta}-D_M\frac{p_S}{p_M}\frac{1+H^2}{n_x}
\frac{\alpha}{\alpha^2+\beta^2}.\label{xx_MIS_var}
\end{eqnarray}

Despite the seemingly simple form of Eqs.\ (\ref{zz_MIS})
and (\ref{xy_MIS}), the right hand sides of those equations do not
constitute explicit expressions for $\lambda$ or for
$\beta/(\alpha^2+\beta^2)$, because those right hand sides
depend upon $\alpha$, $\beta$, $\lambda$ through $n_z$
and $n_x$. Nevertheless,
Eq.\ (\ref{zz_MIS}) does show that $n_z$ must satisfy the following 
inequalities in order for $\lambda$ to be positive:
\be
p_S<n_z<p_S+p_M=1-p_I,\label{MIS_inequalities}
\ee
where the lower bound must satisfy $p_S<p_c=1/3$ and the upper
bound must satisfy $p_S+p_M>p_c=1/3$ (as before, $p_c$ denotes
the percolation threshold, equal to $1/3$ in SEMA) in order
for the entire composite to have a finite, nonzero bulk effective
conductivity.

Eqs.\ (\ref{zz_MIS})--(\ref{xx_MIS_var}) do not admit
closed form solutions. In order to obtain asymptotic
solutions when $|H|\gg 1$, we tried a range of possible ansatzes
for the asymptotic forms of $\alpha$, $\beta$, $\lambda$, and
proceeded to examine them for mathematical consistency and
physical admissibility. Only three of those ansatzes stand up
to both requirements; each is valid for a different
range of the constituent volume fractions $p_M$, $p_I$, $p_S$,
where, of course, $p_M+p_I+p_S=1$. The three ansatzes and
the resulting solutions are described in the following
subsections.

\subsection{The saturating regime}

\subsubsection{General case}

The first ansatz that leads to admissible results is
\be
\alpha\cong\alpha_0,\;\;\;\lambda\cong\lambda_0,\;\;\;
\beta\cong\frac{\beta_0}{H}.\label{ansatz_MIS_sat}
\ee
This leads to the following results:
\be
D_M\cong(1-n_x)^2,\;\;\;D_I\cong(1-n_x)^2\Rightarrow
\frac{D_M}{D_I}\cong 1.
\ee
Consequently, Eq. (\ref{xy_MIS}) yields the following relation
between $\beta_0$ and $\alpha_0$
\be
\beta_0=\alpha_0^2\frac{p_M}{p_I+p_M}.
\ee
Using this to eliminate $\beta_0$ from Eq.\ (\ref{xx_MIS_var}), we get
\begin{eqnarray}
0\cong\frac{1-n_x}{\alpha_0\,n_x\,p_M}(p_S-n_x)\Rightarrow 
n_x\cong p_S<\frac{1}{3}\Rightarrow\nonumber\\
\Rightarrow n_z=1-2n_x\cong1-2p_S=p_M+p_I-p_S>\frac{1}{3}.
\end{eqnarray}
A consequence of the last inequality for $n_z$ is that the 
transformed shape of the spherical inclusions is an {\em oblate spheroid}.
Since we must also have [see Eq.\ (\ref{MIS_inequalities})]
\be
1-2p_S\cong n_z<p_S+p_M=1-p_I,
\ee
we find that we need to require $2p_S>p_I$ in order for this
asymptotic solution to be valid. Eq. (\ref{zz_MIS}) now becomes
\be
\lambda_0=\nu\frac{2p_S(1-3p_S)}{(1-2p_S)(2p_S-p_I)}.
\ee
In order to determine $\alpha_0$ we first need to solve the following
transcendental equation for the eccentricity $e$ of the
oblate spheroid:
\be
n_z=\frac{1+e^2}{e^3}(e-\arctan e)\cong 1-2p_S>\frac{1}{3},
\ee
and then use the relation between $e$ and $\alpha$, $\beta$, $\lambda$
\begin{eqnarray}
e&=&\left(\frac{\alpha^2+\beta^2}{\alpha\lambda}-1\right)^{1/2}\cong
\left(\frac{\alpha_0}{\lambda_0}-1\right)^{1/2}
\equiv e_0\Rightarrow\nonumber\\
&&\;\;\Rightarrow\;\alpha_0=\lambda_0(1+e_0^2).
\end{eqnarray}

These asymptotic results are valid for the range of constituent
compositions defined by
\be
\frac{p_I}{2}<p_S<\frac{1}{3}.\label{range_below}
\ee

When $p_S\rightarrow p_I/2$ from above, $\alpha_0$,
$\lambda_0$, and $\beta_0$ all
diverge, but at different rates:
\begin{eqnarray}
\alpha_0&\propto&\frac{1}{2p_S-p_I},\;\;
\lambda_0=\frac{\alpha_0}{1+e_0^2}\propto\frac{1}{2p_S-p_I},
\label{alpha_lambda_scaling_sat}\\
\beta_0&\propto&\frac{1}{(2p_S-p_I)^2}.\label{beta_scaling_sat}
\end{eqnarray}

It is worth noting that 
$\alpha$ and $\lambda$ are both proportional to $\nu$.
Along with the fact that they are asymptotically independent
of $H$, this indicates that the local electric field in the
$M$ constituent is directed mostly along {\bf B}.  Similar proportionality
is also found in the two-constituent $M/S$ case, discussed below.
It is also worth noting that when $p_S\rightarrow 1/3$ from below,
then also $n_z\rightarrow 1/3$, and consequently $e_0\rightarrow 0$.
Both $\alpha_0$ and $\lambda_0$ tend to 0 as $1-3p_S$, but
$\alpha_0/\lambda_0\rightarrow 1$. This indicates that when the
$S$ constituent approaches its percolation threshold $p_c=1/3$,
the {\em same current flow paths} in the $M$ constituent are
responsible for the leading contribution to the macroscopic
response {\em whatever the direction of the average current
density} $\langle{\bf J}\rangle$.

\subsubsection{Two-constituent $M/S$ mixture}

An important special case of a three-constituent composite with
saturating behavior is $p_I=0$, corresponding to a two-constituent
$M/S$ mixture.  One can also get the results for this case by
setting $p_I=0$ in Eqs.\ (\ref{zz_MIS}), (\ref{xy_MIS})
[or alternatively by taking the limit 
$\rho_1/\rho_2 \rightarrow \infty$ in Eqs.\ (\ref{zz_MM}), (\ref{xy_MM})]
to obtain
the following equations:
\be
\lambda=\nu\left(1-\frac{p_S}{n_z}\right),\;\;\;
\frac{\beta}{\alpha^2+\beta^2}=\frac{H}{1+H^2},\label{zz_xy_MS}
\ee
and noting that Eq.\ (\ref{xx_MIS}) [or Eq.\ (\ref{xx_MM})]
is satisfied as an identity to leading order in $1/H$.

The asymptotic large $|H|$ behavior can
then be obtained by using the ansatz of Eqs.\ (\ref{ansatz_MIS_sat}).
The second of Eqs.\ (\ref{zz_xy_MS}) immediately leads to a simple relation
between $\beta_0$ and $\alpha_0$, namely
\be
\beta_0=\alpha_0^2.
\ee
The value of the ratio $\alpha_0/\lambda_0$
is obtained by first solving the transcendental equation for
the asymptotic eccentricity $e_0$, namely
\be
n_z\cong p_M-p_S=\frac{1+e_0^2}{e_0^3}(e_0-\arctan e_0),
\ee
and then using
\be
e_0\cong\left(\frac{\alpha_0}{\lambda_0}-1\right)^{1/2}.
\ee
Finally, the first of Eqs.\ (\ref{zz_xy_MS}) leads to
\be
\lambda_0\cong\nu\left(1-\frac{p_S}{p_M-p_S}\right)=
\nu\frac{1-3p_S}{p_M-p_S}.
\ee
The fact that
$\alpha_0$, $\beta_0$, $\lambda_0$ all depend upon $\nu$,
i.e., $\alpha_0\propto\nu$, $\lambda_0\propto\nu$,
$\beta_0\propto\nu^2$,
indicates that the leading contribution to the macroscopic response
is due to local currents that flow parallel to {\bf B}
in the $M$ constituent.

When $p_S$ approaches the SEMA percolation threshold value $1/3$
from below, then $n_z\rightarrow 1/3$; hence
$e_0\rightarrow 0$ and $\alpha_0/\lambda_0\rightarrow 1$
from above.
Both $\alpha_0$ and $\lambda_0$ tend to 0 as $1-3p_S$, while
$\beta_0\rightarrow 0$ as $(1-3p_S)^2$. The fact that
$\lambda_0/\alpha_0\rightarrow 1$ again indicates that the
{\em same current flow paths} in the $M$ constituent
are responsible for the leading
contribution to the macroscopic response {\em whatever the
direction of the average current density} $\langle{\bf J}\rangle$.

The SEMA equations for an $M/S$ mixture were treated numerically
in the past. \cite{StroudPRL80} Those calculations are in agreement
with the asymptotic $p_M$ dependence of $\alpha_0$,
$\lambda_0$, and $\beta_0$ obtained for such mixtures in this subsection.

\subsection{The non-saturating regime}

\subsubsection{General case}

Another ansatz which leads to an admissible solution is
\be
\alpha\cong\alpha_0|H|,\;\;\;\lambda\cong\lambda_0|H|,\;\;\;
\beta\cong\beta_0 H.\label{ansatz_MIS_nonsat}
\ee
Because $\lambda$ is now very large, Eq.\ (\ref{zz_MIS})
leads to the following results
\begin{eqnarray}
n_z&\cong&p_S+p_M-{\cal O}\left(1\over|H|\right)=
1-p_I-{\cal O}\left(1\over|H|\right),
\label{nz_MIS_nonsat}\\
n_x&=&\frac{1}{2}(1-n_z)\cong\frac{p_I}{2}+{\cal O}\left(1\over|H|\right).
\end{eqnarray}
Since $p_I<2/3$, we must have $n_z>1/3$ if $|H|$ is large
enough, and the transformed spherical inclusion is then
again an oblate spheroid.

Instead of using the unknowns $\alpha_0$, $\beta_0$, it is convenient
to introduce the variables $x$ and $y$, defined by
\be
x \equiv \frac{\alpha_0^2 + \beta_0^2}{\beta_0} > 1, \;\;\; 
y \equiv \frac{\beta_0}{\alpha_0}. \label{eq:xy_def}
\ee
Using $x$ and $y$, we can write
\begin{eqnarray}
D_M&\cong&\left(1-\frac{p_I}{2}\right)^2
+\frac{p_I^2 y^2}{4}(1-x)^2,\\
D_I&\cong&\left(1-\frac{p_I}{2}\right)^2
+\frac{p_I^2 y^2}{4}.
\end{eqnarray}
Eq.\ (\ref{xy_MIS}) now becomes
\be
(xp_M-p_M-p_I)[(2-p_I)^2+p_I^2 y^2]\cong
p_I^3 y^2(x^2-2x),\label{xy_MIS_1}
\ee
while Eq.\ (\ref{xx_MIS_var}) becomes
\begin{eqnarray}
\lefteqn{
p_Ix(xy^2+1+y^2)\cong}\nonumber\\
&&\;\;\cong\;2p_Ixy^2+2x-\frac{p_S}{p_Mp_I}
[(2-p_I)^2+p_I^2 y^2(1-x)^2].\nonumber\\&&\label{xx_MIS_1}
\end{eqnarray}

These equations are both linear in $y^2$. When $y^2$ is eliminated,
the result is a factorizable cubic equation for $x$
\begin{eqnarray}
0&=&x[x^2(2-3p_I)-x(4+2p_S-5p_I)+2(1+p_S-p_I)]\nonumber\\
&=&x(x-1)\left[x-2\left(1+p_S-p_I\over 2-3p_I\right)\right](2-3p_I).
\end{eqnarray}
Since $p_I=1-p_S-p_M<2/3$, the physical solution is obviously
\be
x=2\frac{1+p_S-p_I}{2-3p_I}=\frac{2(2p_S+p_M)}{2-3p_I}>0.
\ee
This leads to an expression for $y^2$ which is a quotient of
somewhat complicated polynomials in $p_M$, $p_I$, $p_S$. Those
can be factorized, after some effort, leading to
\be
y^2=\frac{(2-3p_I)(2-p_I)(p_I-2p_S)}{p_I^2(2p_S+p_I)},
\ee
which is positive if $p_I>2p_S$. These results lead to the
following expressions for $\alpha_0$, $\beta_0$
\begin{eqnarray}
\beta_0&=&\frac{(2p_S+p_M)(1+p_M+p_S)(p_I/2-p_S)}
{p_I(1-p_I)^2+p_S[2-(2-p_I)^2]},\label{beta_MIS}\\
\alpha_0&=&\left((p_I-2p_S)(2p_S+p_I)(1+p_M+p_S)\over 2-3p_I\right)^{1/2}
\nonumber\\&&\;\times\;
\frac{p_I(p_S+p_M/2)}{p_I(1-p_I)^2+p_S[2-(2-p_I)^2]}.\label{alpha_MIS}
\end{eqnarray}
Obviously, the cubic polynomial which appears in the denominators of
$\alpha_0$, $\beta_0$ is positive over the entire range 
$0<2p_S<p_I$ where Eqs.\ (\ref{beta_MIS}) and (\ref{alpha_MIS})
are applicable. Finally,
$\lambda_0$ can again be found by first solving the following
transcendental equation for the eccentricity $e$ of the oblate
spheroid
\be
n_z=\frac{1+e^2}{e^3}(e-\arctan e)\cong p_S+p_M,
\ee
and then using the relation between $e$ and $\alpha$, $\beta$, $\lambda$
to get
\be
e=\left(\frac{\alpha^2+\beta^2}{\alpha\lambda}-1\right)^{1/2}\cong
\left(\frac{\alpha_0^2+\beta_0^2}{\alpha_0\lambda_0}-1\right)^{1/2}.
\ee

These asymptotic results are valid for the range of constituent
compositions defined by
\be
p_S<\frac{p_I}{2}<\frac{1}{3}.
\ee

When $p_S$ approaches $p_I/2$ from below,
then $\alpha_0$, $\beta_0$, $\lambda_0$ all tend to 0, but at different 
rates:
\begin{eqnarray}
\alpha_0&\propto&\sqrt{p_I-2p_S},\;\;\;\;\lambda_0\cong\frac{\alpha_0}{1+e^2}
\propto\sqrt{p_I-2p_S},\label{critical_above}\\
\beta_0&\propto&p_I-2p_S.\label{critical_above_beta}
\end{eqnarray}

It is worth noting that $\alpha$, $\beta$, and $\lambda$
are all independent of $\nu$. This indicates that the local electric
field in the $M$ constituent has considerable components that
are perpendicular to {\bf B}. It is also worth noting that,
when $p_I\rightarrow 2/3$, both $\alpha_0$ and $\lambda_0$
diverge but $\alpha_0/\lambda_0\rightarrow 1$, i.e.,
\be
\alpha_0\cong\lambda_0\cong2\left(1-9p_S^2\over 3(2-3p_I)\right)^{1/2}.
\label{alpha_lambda_p_I}
\ee
This indicates that, when the total volume fraction of
conducting constituents $p_M+p_S$ approaches its percolation
threshold $p_c=1/3$, the
{\em same current flow paths} in the $M$ constituent
are responsible for the leading
contribution to the macroscopic response {\em whatever the
direction of the average current density} $\langle{\bf J}\rangle$.

\subsubsection{Two-constituent $M/I$ mixture}

\label{MI}

An important special case within the nonsaturating regime is
$p_S = 0$, corresponding to a two-constituent $M/I$ mixture.
One can also get the results for such a mixture by
setting $p_S=0$ in Eqs.\ (\ref{zz_MIS}), (\ref{xy_MIS}), and
(\ref{xx_MIS_var}) to get the following equations:
\begin{eqnarray}
\lambda&=&\nu\frac{1-n_z}{p_M-n_z},\label{zz_MI}\\
\frac{\beta}{\alpha^2+\beta^2}\left(\frac{p_I}{p_M}\frac{D_M}{D_I}+1\right)
&=&\frac{H}{1+H^2},\label{xy_MI}
\end{eqnarray}
\vspace{-0.5 cm}
\be
\frac{\alpha}{n_x}\left(\alpha-\frac{\beta}{H}\right)=
(\alpha^2+\beta^2)\left(1+\frac{\beta}{H}\right)
-2\beta\left(\beta+\frac{\alpha}{H}\right).
\label{xx_MI_var}
\ee

The asymptotic behavior is obtained by making 
the ansatz of Eqs.\ (\ref{ansatz_MIS_nonsat}).
After a sequence of algebraic steps similar to
those described above, the following results are
obtained for the asymptotic linear slopes:
\begin{eqnarray}
\beta_0&=&\frac{1+p_M}{2p_M},\;\;\;\alpha_0=\frac{p_I}{2p_M}
\left(1+p_M\over 3p_M-1\right)^{1/2},\label{beta_alpha_MI}\\
\lambda_0&=&\alpha_0\frac{(2p_M/p_I)^2}{1+e_0^2},\label{lambda_MI}
\end{eqnarray}
where $e_0$ is given implicitly in terms of $p_M$ by the transcendental
equation
\be
p_M=\frac{1+e_0^2}{e_0^3}(e_0-\arctan e_0),\label{pM_MI}
\ee
which must be solved numerically. As in the $M/I/S$
nonsaturating regime, $\alpha_0$, $\beta_0$, $\lambda_0$ are all 
independent of $\nu$. This indicates that the leading
contribution to the macroscopic or bulk effective response
is due to local currents that flow perpendicular to {\bf B}
in the $M$ constituent.

When $p_M$ approaches the SEMA percolation threshold value of $1/3$
from above, then $\beta_0\rightarrow 2$ and
$e_0\rightarrow 0$. Therefore $\lambda_0/\alpha_0\rightarrow 1$
from above, and both $\alpha_0$ and
$\lambda_0$ diverge as $1/\sqrt{3p_M-1}$. The fact that
$\lambda_0/\alpha_0\rightarrow 1$ again indicates that the
{\em same current flow paths} in the $M$ constituent
are responsible for the leading
contribution to the macroscopic response {\em whatever the
direction of the average current density} $\langle{\bf J}\rangle$.

Once again, in order to confirm the asymptotic behavior predicted
analytically, we have solved the SEMA equations numerically.  We assume
$\nu_1\equiv\nu = 1$ [see Eq.\ (\ref{rho_1_rho_2})],
$\hat{\rho}_2 =\infty$, and consider a variety
of values of $p_1\equiv p_M$ above the percolation threshold $p_c$
(equal to 1/3 in the SEMA
in three dimensions).  The resulting behavior of $\alpha$ and $\lambda$ 
is shown in Figs.\ \ref{magnetoins}(a) and \ref{magnetoins}(b); 
we also show the relative Hall resistivity $\beta/H$
in Fig.\ \ref{magnetoins}(c).  
Evidently, for any choice of $p_M$, 
$\alpha$ and $\lambda$ rapidly approach their asymptotic linear dependence
on $|H|$, as predicted by the asymptotic analysis.  Furthermore, the slope
increases as $p_M$ approaches the percolation threshold, again as predicted
by the asymptotic results.   The asymptotic linear dependence appears to
be reached approximately when $|H| > 5$, for all the values of $p_M$
that were considered.
For $p_M$ less than about $0.5$, we had
some difficulty, using our simple algorithm, in solving
the SEMA equations numerically; by contrast, of course, the asymptotic
analysis gives the slope, for any value of $p_M$ greater than 
$p_c = 1/3$, without any difficulty.

\subsection{Scaling behavior near the transition point}
\label{MIS_scaling}

If we compare the critical behaviors exhibited by
$\alpha$, $\beta$, $\lambda$ when $|H|\gg 1$ and $p_I\rightarrow 2p_S$
from above [Eqs.\ (\ref{critical_above}), (\ref{critical_above_beta})]
and from below [Eqs.\ (\ref{alpha_lambda_scaling_sat})
and (\ref{beta_scaling_sat})], we are led to anticipate a scaling
behavior.  That is, we should be able to describe the critical behavior in
both regimes by using a scaling variable which is
some power of $H^2(p_I-2p_S)^3/\nu^2$, and writing
$\alpha$, $\beta$, $\lambda$ in terms of three (scaling) functions
of that variable. We have found that the most convenient scaling
variable for this purpose is
\be
Z\equiv\left(\frac{|H|}{\nu}\right)^{2/3}(p_I-2p_S).\label{Z_def}
\ee
In the ``critical region'', i.e., when $|H|\gg 1$ and $|p_I-2p_S|\ll 1$,
the bulk effective resistivity parameters can now be expressed as
\begin{eqnarray}
\alpha&\cong&\frac{\nu}{p_I-2p_S}F_\alpha(Z),\\
\lambda&\cong&\frac{\nu}{p_I-2p_S}F_\lambda(Z),\\
\beta&\cong&\frac{\nu^2}{H(p_I-2p_S)^2}F_\beta(Z).
\end{eqnarray}

\begin{figure}[H]
\centerline{
\epsfig{figure=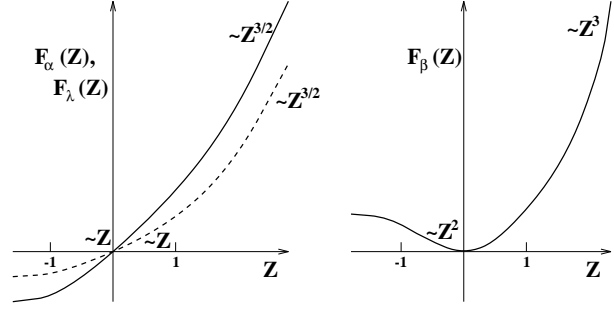,height=4 cm}
}
\vspace{1 cm}
\caption{Qualitative plots of the scaling functions obtained using
SEMA. The left
plot shows $F_\alpha(Z)$ (solid line) and $F_\lambda(Z)$ (dashed
line), which are similar
but not identical, with $|F_\alpha(Z)|>|F_\lambda(Z)|$, 
while the right plot shows $F_\beta(Z)$, which is never negative.
}
\label{scaling_functions}
\end{figure}

As usual, there are three important extreme regimes within the
critical region, namely: $Z < 0$, $|Z| \gg 1$ (Regime I, where $2p_S>p_I$);
$Z > 0$, $|Z| \gg 1$ (Regime II, where $2p_S<p_I$); and
$|Z|\ll 1$ (Regime III, where we can have either $2p_S>p_I$,
or $2p_S<p_I$, or $2p_S=p_I$). 
The behavior of the scaling
functions $F_\alpha(Z)$, $F_\lambda(Z)$, $F_\beta(Z)$ in
Regimes I and II is dictated by the critical behaviors
found earlier. Their behavior in Regime III is dictated
by the requirement that this behavior must exactly 
cancel the divergences that
would otherwise occur due to the vanishing $2p_S-p_I$
factor in the denominators of the above expressions.
These considerations lead to the following forms for the
scaling functions in the three regimes:
\begin{eqnarray}
F_\alpha(Z)&\cong&\left\{\begin{array}{ll}
-A & \mbox{Regime I}\\
A'Z^{3/2} & \mbox{Regime II}\\
A''Z & \mbox{Regime III}\end{array}\right.\\
F_\lambda(Z)&\cong&\left\{\begin{array}{ll}
-L & \mbox{Regime I}\\
L'Z^{3/2} & \mbox{Regime II}\\
L''Z & \mbox{Regime III}\end{array}\right.\\
F_\beta(Z)&\cong&\left\{\begin{array}{ll}
B & \mbox{Regime I}\\
B'Z^3 & \mbox{Regime II}\\
B''Z^2 & \mbox{Regime III}\end{array}\right.
\end{eqnarray}
where the primed and double primed and unprimed versions of
$A$, $B$, $L$ are positive dimensionless constants of order 1.
Their values can be found, if necessary, by comparing the
resulting expressions for $\alpha$, $\beta$, $\lambda$ with
the detailed solutions of the SEMA equations. Qualitative
plots of these scaling functions are shown in Fig.\
\ref{scaling_functions}.

As indicated above, these equations make a nontrivial prediction
about the behavior
of $\alpha$, $\beta$, $\lambda$ in Regime III, which was
not worked out in the previous sections, namely:
\be
\alpha\cong\alpha_0|H|^{2/3},\;\;\;\lambda\cong\lambda_0|H|^{2/3},\;\;\;
\beta\cong\beta_0|H|^{1/3}\,{\rm sign}(H).\label{ansatz_crit}
\ee
We now use these scaling forms as an ansatz for another asymptotic solution
of the SEMA equations, assuming $p_I=2p_S$,
in order to check for consistency of
the scaling scheme developed here.

Since $\lambda$ is very large, Eq.\ (\ref{zz_MIS}) entails
\be
n_z\cong p_S+p_M=1-p_I=1-2p_S>\frac{1}{3},\;\;n_x\cong p_S<\frac{1}{3},
\ee
where we have used the fact that $p_I=2p_S$. From Eq.\ (\ref{zz_MIS})
we can also calculate the small correction to $n_z\cong 1-2p_S$ in
terms of $\lambda_0$
\be
1-2p_S-n_z\cong\frac{\nu}{\lambda_0|H|^{2/3}}\frac{2p_S(1-3p_S)}{1-2p_S}.
\label{nz_diff}
\ee
In addition, we can write expressions for $D_M$, $D_I$, and for their ratio
$D_M/D_I$, that go {\em beyond} leading order in $1/H$, namely
\begin{eqnarray}
D_M&\cong&(1-n_x)^2+\frac{n_x^2\beta_0^2}{\alpha_0^2|H|^{2/3}}
\left(1-\frac{\alpha_0^2}{\beta_0}\right)^2,\\
D_I&\cong&(1-n_x)^2+\frac{n_x^2\beta_0^2}{\alpha_0^2|H|^{2/3}},\\
\frac{D_M}{D_I}&\cong& 1+\frac{\alpha_0^2-2\beta_0}{|H|^{2/3}}
\left(n_x\over 1-n_x\right)^2.
\end{eqnarray}

  From Eq.\ (\ref{xy_MIS}) we get a relation between $\beta_0$
and $\alpha_0$
\be
\beta_0=\alpha_0^2\frac{1-3p_S}{1-p_S}.
\label{beta_0_crit}
\ee
Using this result, we find that Eq.\ (\ref{xx_MIS}) becomes an
identity to leading order in $1/H$.
We therefore need to consider that
equation in the next-to-leading order. The result is
\be
1-2p_S-n_z\cong\frac{\alpha_0^2}{|H|^{2/3}}
\frac{4p_S^2(1-3p_S)}{(1-p_S)^3}.
\ee
Comparison of this equation with Eq.\ (\ref{nz_diff}) yields the
following relation between $\alpha_0$ and $\lambda_0$:
\be
\frac{\alpha_0^2\lambda_0}{\nu}=\frac{(1-p_S)^3}{2p_S(1-2p_S)}.
\label{al_square_lam_crit}
\ee
Another relation between those two unknowns is obtained by
first solving the transcendental equation for the eccentricity $e$
of the oblate spheroid
\be
n_z=\frac{1+e^2}{e^3}(e-\arctan e)\cong p_S+p_M,
\ee
and then using the expression for $e$ in terms of
$\alpha$, $\beta$, $\lambda$ to get
\be
1+e^2=\frac{\alpha^2+\beta^2}{\alpha\lambda}\cong\frac{\alpha_0}{\lambda_0}.
\label{al_over_lam_MIS_crit}
\ee
Eqs. (\ref{beta_0_crit}), 
(\ref{al_square_lam_crit}), and
\ (\ref{al_over_lam_MIS_crit})
then provide a complete and consistent solution
for $\alpha_0$, $\beta_0$, $\lambda_0$, obtained by jointly considering
the scaling and SEMA equations.

Again, it is worth noting that
both $\alpha$ and $\lambda$ are proportional to $\nu^{1/3}$,
while $\beta\propto\nu^{2/3}$.
This seems to be consistent with the fact that $\alpha$
and $\lambda$ are also proportional to $|H|^{2/3}$,
while $\beta\propto|H|^{4/3}/H$. These behaviors indicate that the
local current flows both parallel and perpendicular to
{\bf B}, and that the three principal Ohmic conductivities of
the $M$ constituent all contribute to the macroscopic
response when $p_I=2p_S$.

\section{Summary and discussion}
\label{summary}

A striking result of the present work is that, in an $M/I$ composite, 
both $\rho^{(e)}_\perp$ and $\rho^{(e)}_\parallel$
are proportional to $|H|$ in the 
strong-field limit.   Such {\em linear magnetoresistance} has long been
a mysterious observed feature of transport in polycrystalline samples of
even so-called ``simple metals''. \cite{Linear}
This behavior has sometimes been ascribed to 
macroscopic inhomogeneities.  But until
now, only in the low-concentration limit has proof been given
that such inhomogeneities could actually produce a linear 
magnetoresistance. \cite{StroudPanPRB76,SampsellGarlandPRB76,Balagurov86} 
Here, we have shown that
both $\rho^{(e)}_\perp$ and $\rho^{(e)}_\parallel$
remain linear in $|H|$ even at
{\em higher concentrations} of inclusions, provided those inclusions have
{\em strictly zero conductivity}.  This result may be relevant to
a range of experimental systems.

It is quite easy to understand, qualitatively, the behavior of
an $M/S$ disordered mixture: Whatever the direction of the
{\em average current density} $\langle{\bf J}\rangle$, the
{\em local current density} ${\bf J}({\bf r})$ in the $M$
constituent will flow mostly along ${\bf B}\parallel z$
when $H$ is very large, and this tendency will become more
and more pronounced with increasing $|H|$. Components of
${\bf J}({\bf r})$ that are perpendicular to ${\bf B}$
will have finite values only inside the $S$ constituent,
while in the $M$ constituent they will tend to 0 as $1/H^2$.
This results in a current flow pattern that saturates when
$|H|\gg 1$; therefore, all the Ohmic resistivities will also
saturate. The Hall resistivity will be very small---of order
$1/H$---because the local Hall field ${\bf E}_H$ will also
be of that order.

The behavior of $M_1/M_2$ mixtures is more difficult to
understand qualitatively. Our interpretation of the
(successful) ansatz of Eq.\ (\ref{ansatz_MM}) is that
when $\langle{\bf J}\rangle\parallel{\bf B}\parallel z$,
the current distribution saturates for large values of $|H|$. 
But when $\langle{\bf J}\rangle\perp{\bf B}$, then that distribution
continues to evolve with increasing $|H|$, with the current
distortions increasing asymptotically as $|H|^{1/3}$.

The macroscopic response of $M/I$ mixtures is qualitatively
similar to the behavior of an isolated insulating inclusion
embedded in an $M$ host, as shown many years ago by an exact
solution for such an inclusion of spherical or cylindrical shape.
\cite{SampsellGarlandPRB76}
The fact that this kind of behavior persists down to the percolation
threshold indicates that, despite the interactions between
current distortions produced by different inclusions, the
local current distribution never saturates as $|H|$ increases.
The results we got would require that the current distortions
increase as $|H|^{1/2}$ for large $H$. However, the fact that
the coefficients $\alpha_0$ and $\lambda_0$ diverge as
$p_M\rightarrow p_c$ probably signals that those distortions
increase even more rapidly than $|H|^{1/2}$ precisely at the percolation
threshold.

The present results can also
be compared with some simulations performed previously on a
discrete network
model, where finite size $L\times L\times L$ samples were
considered precisely at the percolation threshold.
\cite{SarBergStrelPRB93}
Those simulations showed that, when $L$ is much less than a 
``magnetic correlation length'' $\xi_H\propto |H|^{0.46}$,
the Ohmic resistivities saturate at a value proportional to
$L^{6.7}$. However, in the opposite limit $L\gg\xi_H$, the
Ohmic resistivities continue to increase as $|H|^{2.1}L^{2.2}$.
This result is consistent, within the error bars, 
with $H^2L^{t/\nu}$, where $t\cong 2.0$ and
$\nu\cong 0.89$ are the usual percolation critical exponents
for scalar Ohmic conductivity ($\sigma_e\propto\Delta p^t$,
$\Delta p\equiv p_M-p_c$) and for the percolation correlation length
($\xi_p\propto\Delta p^{-\nu}$)---see Ref.\
\onlinecite{AharonyStauffer92}. Using
finite size scaling to replace the system size $L$ by
$\xi_p$ in these results, we are led to expect that
\be
\alpha,\lambda\propto
\left\{\begin{array}{lll}
\Delta p^{-4.0} & {\rm for} & \Delta p\gg|H|^{-0.52}\;{\rm or}\;
\xi_p\ll\xi_H,\\
H^2 & {\rm for} & \Delta p\ll|H|^{-0.52}\;{\rm or}\;
\xi_p\gg\xi_H.\end{array}
\right.\label{simulations}
\ee
Obviously, this behavior differs in a number of ways from
what was found in Section \ref{MI} using SEMA.
It is not surprising that the critical exponents predicted by
SEMA are quantitatively incorrect. However, the fact 
that, according to the simulations, the magnetoresistivity
saturates as $|H|\rightarrow\infty$, is
qualitatively at odds with the predictions of SEMA.
Clearly, this qualitative discrepancy needs to be examined further.
We conjecture that it has to do with the existence of three
diverging lengths in this problem, namely $\xi_p$, $\xi_H$,
and $L$.

The qualitative situations described above continue to be
applicable also in the case of the three-constituent
$M/I/S$ mixtures. The presence or
absence of system-spanning (i.e., infinite) 
current flow paths, which are parallel
to {\bf B} inside the $M$ constituent, now depends on the relative
amounts of $I$ and $S$ inclusions. The critical points or transition
points $p_I=2p_S$ can be expected to occur when such paths
first appear with increasing $p_S$.

Obviously, this transition is a kind of percolation process. In fact,
we believe it is a physical realization of ``anisotropic
percolation''. This kind of percolation
was first considered many years ago
in the context of a two-dimensional, randomly diluted,
square bond network. \cite{SykesEssamPRL63}
The geometrical properties of anisotropic percolation in
hypercubic random bond networks of arbitrary dimension were
also studied extensively. \cite{RednerStanley79,Nakanishi81}

As originally defined, the anisotropic percolation problem
is characterized by different occupation probabilities for
bonds aligned along different principal axes of the network.
This situation is not easily implemented in a continuum percolating 
system, because that would require an anisotropic constituent
where the principal conductivities have ratios that are
extremely different from 1. Also, the principal axes of different grains of
that constituent would have to be {\em identically oriented}.
While this may be difficult to achieve in conducting materials
when ${\bf B}=0$, such extreme ratios and identical orientations
can easily be attained, even in an isotropic
conductor, just by applying a magnetic field such that
$|H|\gg 1$ or $\nu\ll 1$.
In that case, if we identify volume fractions with the
bond occupation probabilities, then we can say that, for
$|H|\gg 1$ or $\nu\ll 1$, the regions that are highly conducting along
{\bf B} occupy a fraction $p_M+p_S$ of the total volume. By contrast,
the regions that are highly conducting perpendicular to
{\bf B} occupy the smaller volume fraction $p_S$. Our SEMA
result, for the transition points between saturating and
non-saturating regimes of magnetoresistance,
can then be interpreted as follows
\begin{eqnarray}
0&=&1-p_x-p_y-p_z=1-p_S-p_S-(p_M+p_S)\nonumber\\
&=&p_I-2p_S,
\end{eqnarray}
where $p_z=p_M+p_S$ represents the bond occupation probability along
{\bf B} while $p_x=p_y=p_S$ represent the bond occupation probabilities
in the directions perpendicular to {\bf B}.

The identification of the critical points in the $M/I/S$
magnetoresistive response with an anisotropic percolation 
threshold needs to be verified by a more accurate treatment
of the bulk effective magnetoresistance. In particular, it
needs to be determined whether the ever-present Hall
conductivity is an irrelevant perturbation, or whether
it in fact destabilizes the usual percolation fixed point,
or alters the critical behavior associated with it. In any case, we expect 
that such a treatment will yield different values for the
critical exponents and different forms for the scaling functions
and scaling variable than those obtained here using SEMA.

Experimental studies of the line of magnetoresistive critical points
can be conducted on samples made by using a semiconductor host
with a large Hall mobility $\mu$ as the $M$ constituent, in
which two types of inclusions are randomly embedded: highly insulating
inclusions (e.g., voids) as the $I$ constituent, and either
highly conducting normal metallic inclusions
or superconducting inclusions as the $S$ constituent.

\acknowledgements

\begin{figure}[H]
\centerline{\epsfig{figure=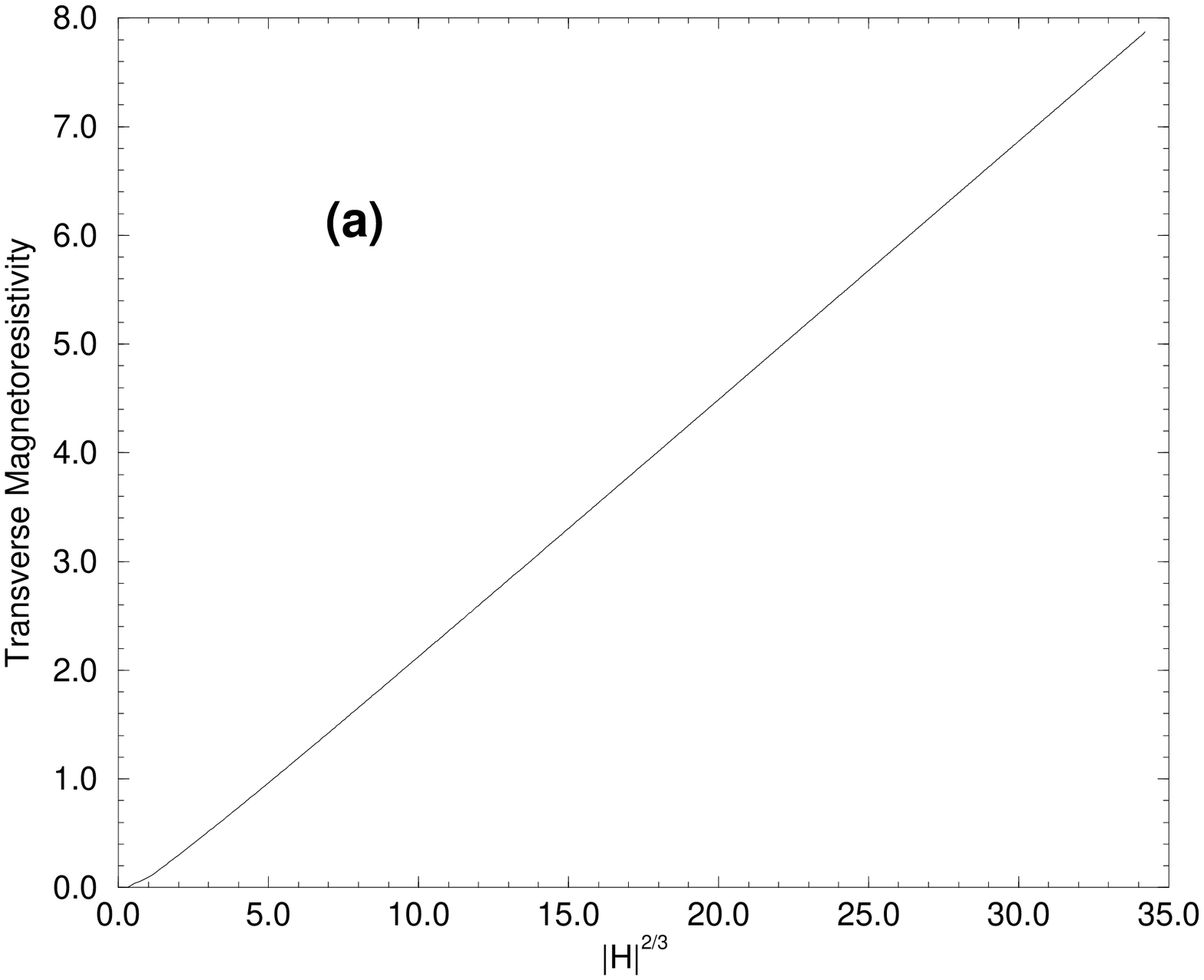,height=9.5 cm,angle=-90}
\epsfig{figure=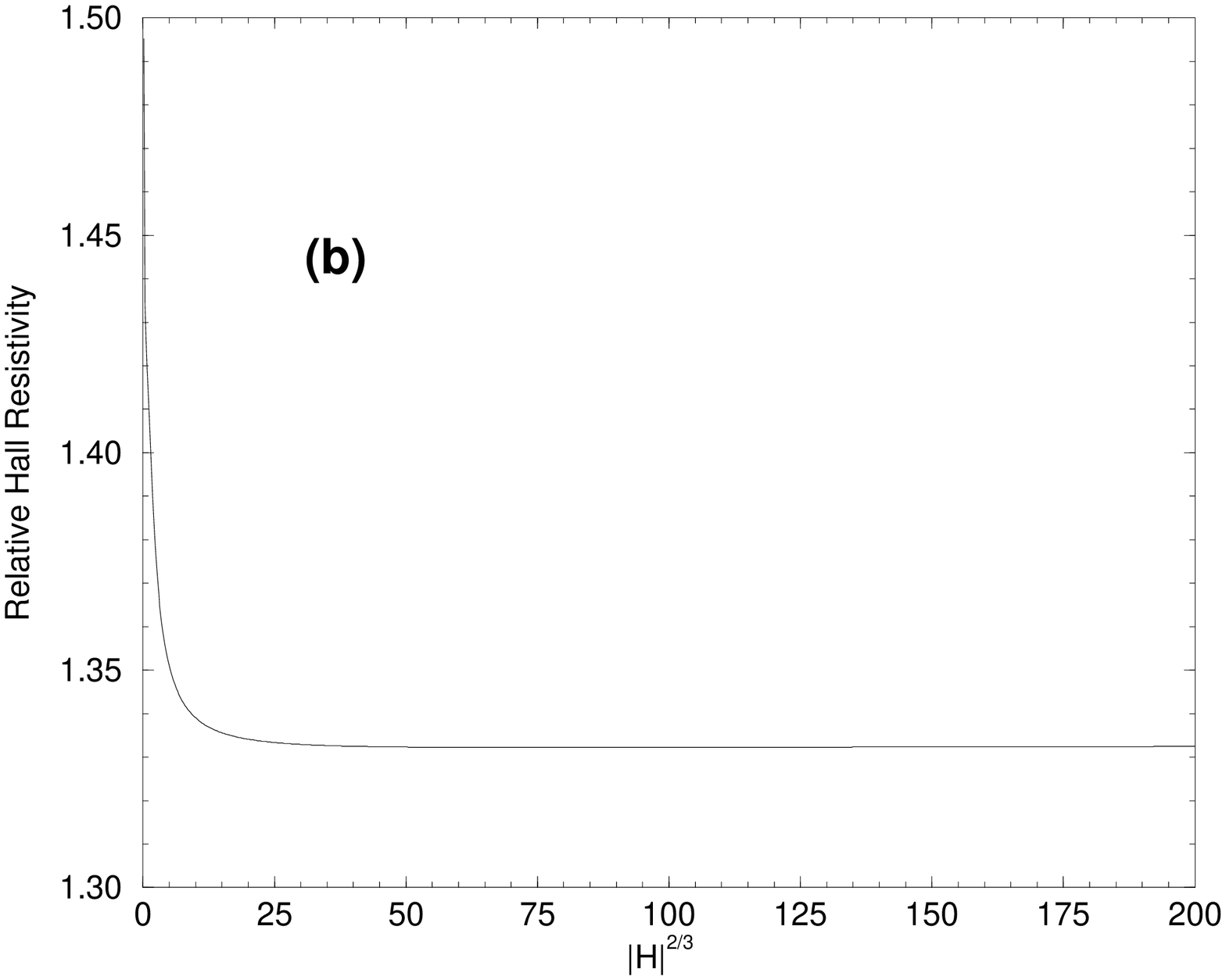,height=9.5 cm,angle=-90}
}
\vspace{1 cm}
\caption{Results of numerical solution of the SEMA Eqs.\
(\ref{zz_MM})--(\ref{xx_MM}) for an $M_1/M_2$ mixture.
(a) Plot of the relative bulk effective
transverse magnetoresistivity $\alpha-1$ vs.\ $|H|^{2/3}$ for
a three-dimensional composite of two free-electron-metal
constituents, present in equal amounts $p_1=p_2=1/2$,
which have the same ohmic resistivity,
but Hall resistivities that differ by a factor two [i.e.,
$\rho_1=\rho_2$ and $H_1 = H$,
$H_2 = 2H$, in the notation of Eq.\ (\ref{rho_1_rho_2})].
(b) Same as (a), except
that we plot the relative bulk effective Hall resistivity
$\beta(H)/H$ vs.\ $|H|^{2/3}$ [cf.\ Eq.\ (\ref{rho_e})].
}
\label{magnetocond}
\end{figure}

\begin{figure}[H]
\centerline{\epsfig{figure=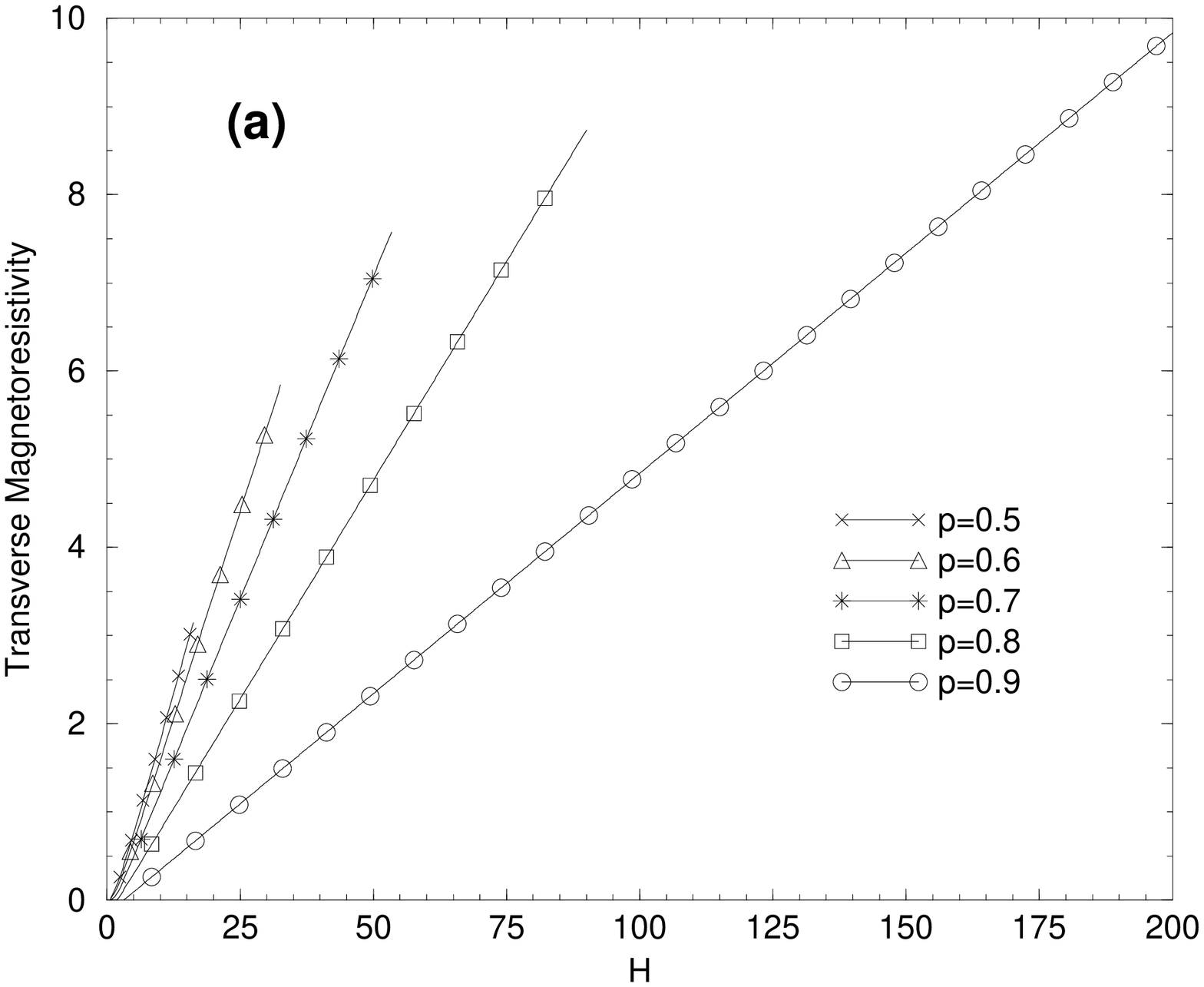,height=9.5 cm,angle=-90}
\epsfig{figure=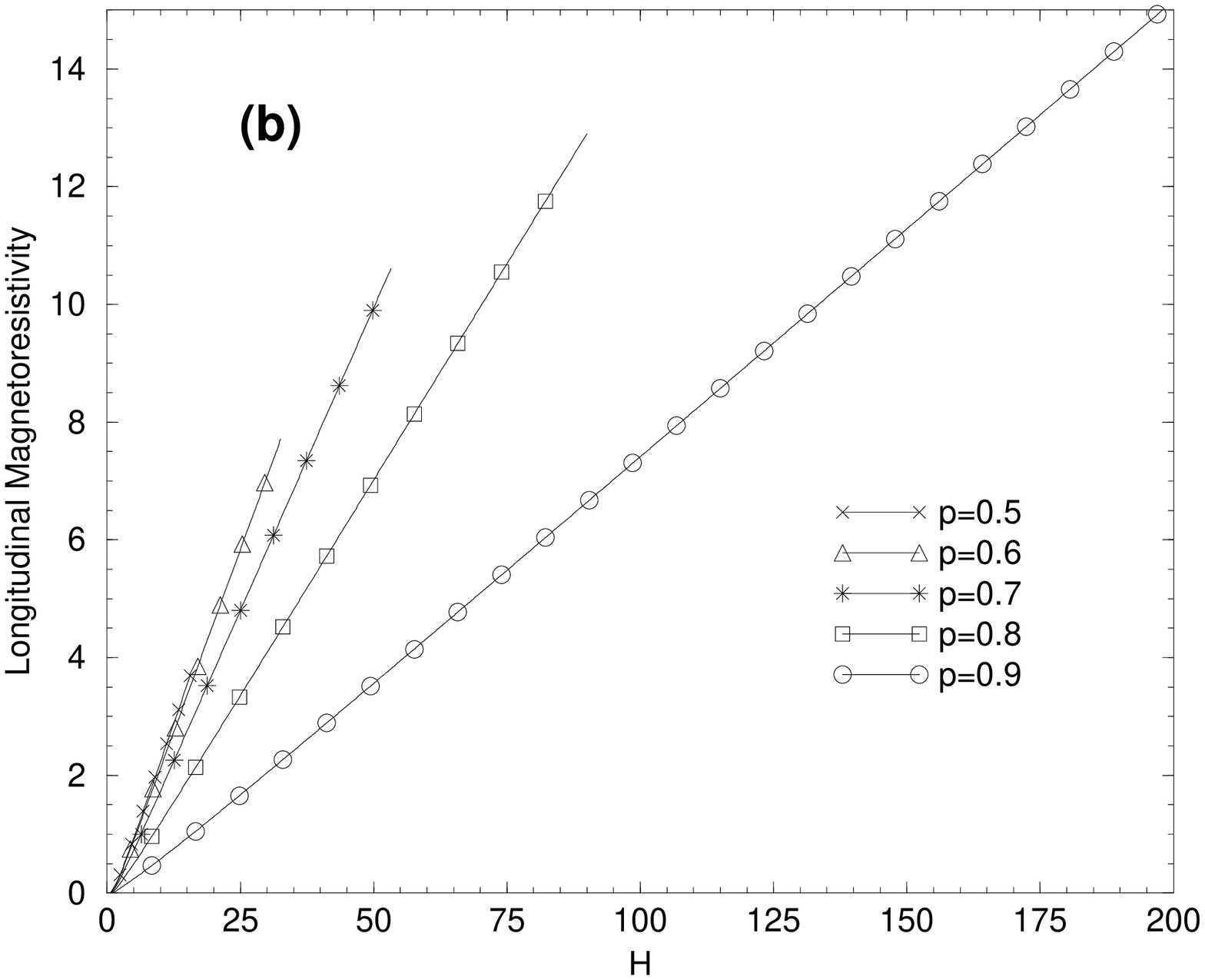,height=9.5 cm,angle=-90}
}
\epsfig{figure=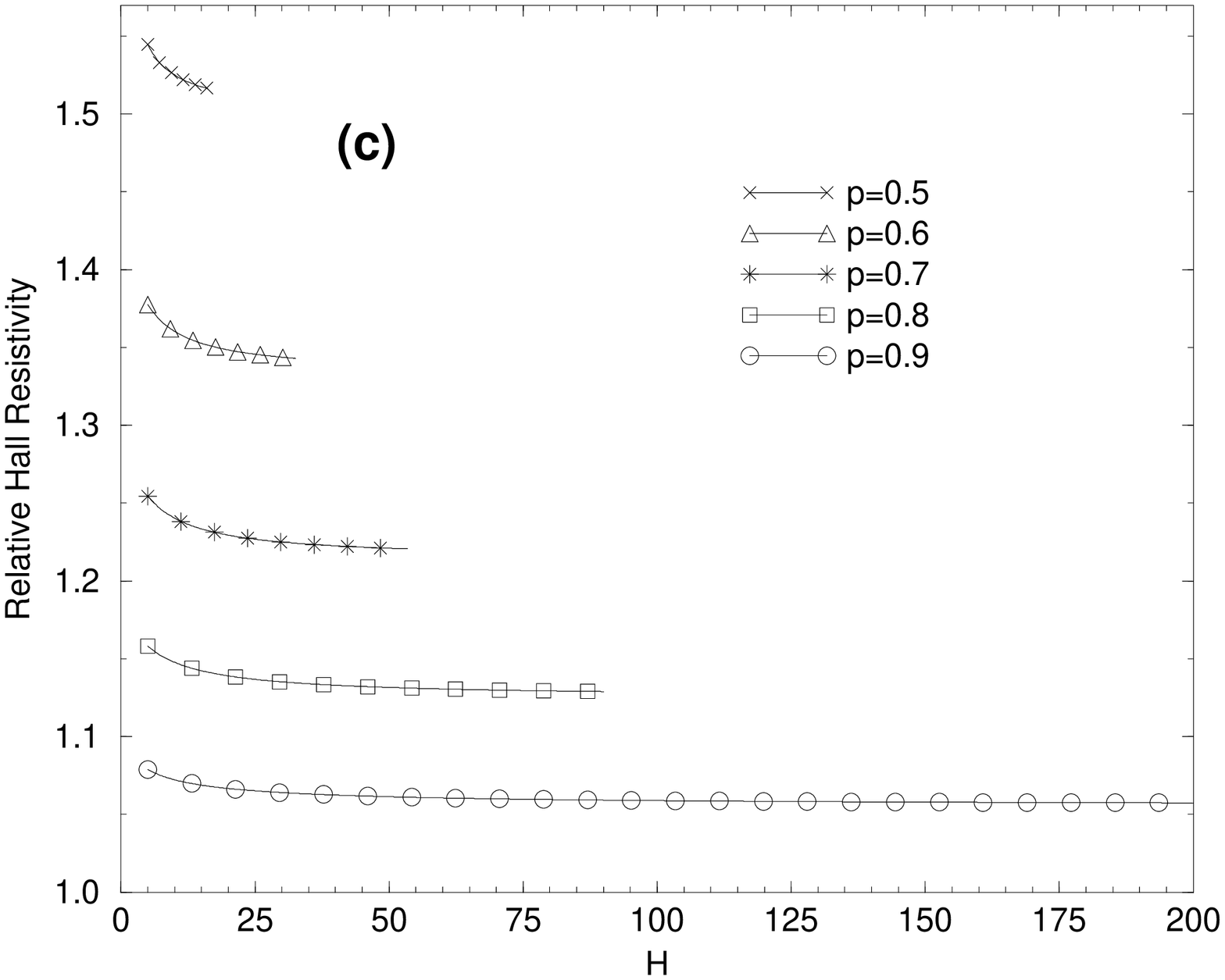,height=9.5 cm,angle=-90}
\vspace{1 cm}
\caption{Results of numerical solution of the SEMA Eqs.\
(\ref{zz_MI})--(\ref{xx_MI_var}) for an $M/I$ mixture.
(a) Plot of the relative bulk effective
transverse magnetoresistivity $\alpha-1$ vs.\ $|H|$ for
a three-dimensional two-constituent composite of free-electron
metal and perfect insulator for various values of the metal
volume fraction $p \equiv p_M$.
(b) Same as (a), except that we plot the relative bulk effective
longitudinal magnetoresistivity $\lambda-1$.
(c) Same as (a), except that we plot the relative
bulk effective Hall resistivity $\beta(H)/H$ vs.\ $|H|$
only at high fields ($|H| > 5$).
}
\label{magnetoins}
\end{figure}

We are grateful to Eivind Almaas for help with Figs.\
\ref{magnetocond} and \ref{magnetoins}.
This research was supported in part by grants from the
US-Israel Binational Science Foundation, the Israel Science Foundation,
and NSF Grant DMR 97-31511.


\widetext

\end{document}